\documentstyle[mncite,psfig]{mn}

\def\gte{\,\lower.6ex\hbox{$\buildrel >\over \sim$} \, }
\def\lte{\,\lower.6ex\hbox{$\buildrel <\over \sim$} \, }
\ifCUPmtlplainloaded \else

\fi
\newcommand{\Htwo}{\mbox{H$_2$}}
\newcommand{\Hyd}{{\rm H}}
\newcommand{\K}{{\rm K}}

\title{In-Shock Cooling in Numerical Simulations}

\author[Roger M. Hutchings and Peter A. Thomas]{Roger
            M. Hutchings\thanks{Email: rogerh@astr.cpes.susx.ac.uk} and
            Peter A. Thomas\\  
            Astronomy Centre, University of Sussex, 
            Falmer, Brighton BN1 9QJ, UK.}

\date{}

\begin{document}

\maketitle

\begin{abstract}
We model a one-dimensional shock-tube using smoothed particle
hydrodynamics and investigate the consequences of having finite
shock-width in numerical simulations caused by finite resolution of
the codes. We investigate the cooling of gas during passage through
the shock for three different cooling regimes.

For a theoretical shock temperature of 10$^5$K, the maximum
temperature of the gas is much reduced.  When the ration of the
cooling time to shock-crossing time was 8, we found a reduction of 25
percent in the maximum temperature reached by the gas.  When the ratio
was reduced to 1.2 maximum temperature reached dropped to
50 percent of the theoretical value.  In both cases the cooling time
was reduced by a factor of 2.

At lower temperatures, we are especially interested in the production
of molecular Hydrogen and so we follow the ionization level and \Htwo\
abundance across the shock.  This regime is particularly relevent to
simulations of primordial galaxy formation for halos in which the virial
temperature of the galaxy is sufficiently high to partially re-ionize
the gas.  The effect of in-shock cooling is substantial: the
maximum temperature the gas reaches compared to the theoretical
temperature was found to vary between 0.15 and 0.81 for the
simulations performed, depending upon the strength of the shock and
the mass resolution.   The downstream ionization level is
reduced from the theoretical level by a factor of between $2.4$
and $12.5$, and the resulting \Htwo\ abundance was found to be reduced
to a fraction of $0.45$ to $0.74$ of its theoretical value.

At temperatures above 10$^5$K, radiative shocks are unstable and will
oscillate.  We reproduce these oscillations and find good agreement
with the previous work of Chevalier and Imamura (1982), and Imamura,
Wolff and Durisen (1984). The effect of in-shock cooling in such
shocks is difficult to quantify, but is undoubtedly present.

We conclude that extreme caution must be exercised when interpretting
the results of simulations of galaxy formation.
    
\end{abstract}


\section{Introduction}
In order for hydrodynamics codes to be able to simulate shocks it is
necessary that the width of the shock be spread over a few mesh cells,
or represent a few inter-particle spacings, depending on whether a
grid or particle based code is being used. In reality, however, the true width
of the shock is only a few mean free paths, giving a shock thickness
$\Delta x$, of
\begin{equation}
\Delta x\sim \frac{1}{n\sigma}
\end{equation}
where $n$ is the number density and $\sigma$ is the collisional
cross-section.
A few simple calculations show that the true shock thickness is orders
of magnitude smaller than the shock thickness obtained in 
simulations. This allows the possibility that gas in simulations can
cool artificially while passing through a shock, and that temperature-sensitive
quantities, such as ionization levels and molecular abundances may
also vary. Any such effects will be purely numerical because of the
finite shock crossing time in simulations.
In the present paper will consider a number of possibilities. Firstly,
if gas is able to cool during a shock the maximum temperature the gas
will reach after
passing through the shock will be reduced from that of an adiabatic
jump.
Where the cooling function rate decreases with
increasing temperature,
this will then reduce the cooling time of the post-shock gas.
Secondly, we shall consider the evolution of the ionization level
through a shock, where the cooling rate is proportional to the number
density of free electrons. This has important consequences for
simulations of primordial structure formation as the ionization level
determines the rate of molecular hydrogen (\Htwo) and Hydrogen-Deuterium (HD)production. These two molecules then determine the cooling time of the
gas down to temperatures less than $100$K.
Finally, we shall also address the question of instability of radiative shocks.
Imamura et al. (1984) and Chevalier and Imamura (1982) found that
radiative shocks with a cooling rate per unit volume $\propto
\rho^{2}T^{\alpha}$ (where $\rho$ is the mass density and $T$ the temperature) are unstable against perturbations when $\alpha
\leq 0.4$. This instability results in periodic 
oscillations of the position of the shock front and of the jump
temperature at the shock front. We address the nature of these
oscillations in the context of galaxy formation using a cooling
function composed of the sum of a number of terms, each with a
different power law.

\section{Analytic Profiles}
Shocks can be generally classified into three types,
depending upon the amount of cooling present. In adiabatic shocks the
shocked gas does not cool significantly 
during the period of time of interest. Isothermal shocks have the same
jump conditions at the shock front as an adiabatic shock, but the
cooling is very rapid so that the downstream gas is at the
same temperature as the upstream gas. Any cooling tail is thus very narrow.
In between these two extremes is the case where the shocked gas has a
cooling time of order the time-scale which is of interest in a given
situation.

For an adiabatic shock the conservative form of the fluid equations in
the absence of sources which change momentum, mass or energy can be
written as
\begin{equation}
\frac{d}{dx}\left(\rho u\right)=0
\\
\label{equ:if1}
\end{equation}
\begin{equation}
\frac{d}{dx}\left(\rho u^{2}+P\right)=0
\\
\label{equ:if2}
\end{equation}
\begin{equation}
\frac{d}{dx}\left[\rho
u\left(\frac{1}{2}u^{2}+\epsilon+\frac{P}{\rho}\right)\right]=0
\label{equ:if3}
\end{equation}
where the variables
$u$, $x$, $\rho$, $\epsilon$ and $P$ 
represent velocity, position, density, specific energy and
pressure respectively, all evaluated in the rest frame of
the shock. 

If we apply equations \ref{equ:if1}-\ref{equ:if3} to gas far upstream
and write its properties in terms of gas far downstream, we arrive at
the Rankine-Hugoniot jump conditions:
\begin{equation}
\rho_{2}u_{2}^{2}+P_{2}=\rho_{1}u_{1}^{2}+P_{1}
\\
\end{equation}
\begin{equation}
\rho_{2}u_{2}=\rho_{1}u_{1}
\\
\end{equation}
\begin{equation}
\frac{1}{2}u_{2}^{2}+\frac{\gamma}{\gamma-1}k_{b}\frac{T_{2}}{\mu
m_\Hyd}=\frac{1}{2}u_{1}^{2}+\frac{\gamma}{\gamma-1}k_{b}\frac{T_{1}}{\mu
m_\Hyd}
\label{equ:adi}
\end{equation}
where $T$ is the gas temperature, $k_{b}$ the Boltzmann constant $\mu$
the mean molecular weight and $m_\Hyd$ the mass of a hydrogen
atom. The suffixes $1$ and $2$ refer to the upstream and downstream
gas respectively and the equation of state of the gas is given by the
equations:
\begin{equation}
\frac{P}{\rho}=\frac{k_{b}T}{\mu m_\Hyd}
\end{equation}
\begin{equation}
\epsilon=\frac{1}{\gamma -1}\frac{k_{b}T}{\mu m_\Hyd}
\end{equation}
Taking $\gamma=5/3$ we can write the pressure of the shocked
gas as 
\begin{equation}
P_{2}=\frac{P_{1}\left(4\rho_{2}-\rho_{1}\right)}
{\left(4\rho_{1}-\rho_{2}\right)}
\label{equ:stop}
\end{equation}

For the case of an isothermal shock where the
temperature of the gas far downstream is equal to its temperature up
stream, the conservation of energy equation is no
longer valid and should be replaced with the condition $T_{1}=T_{2}$.
We must stress however that the gas will still undergo
an adiabatic jump at the shock front before cooling back down to its
upstream temperature.

For the case of a radiative shock we wish to determine the path
between the state of the gas at the shock front and its state
downstream. In order to do this we must add a cooling term to the
energy flux equation:
\begin{equation}
\frac{d}{dx}\left[\rho
u\left(\frac{1}{2}u^{2}+\epsilon+\frac{P}{\rho}\right)-\rho^{2}\Lambda
\right]=0
\\
\label{equ:apf3}
\end{equation}
The cooling term $\Lambda$ serves to determine the path between the
down stream and adiabatic jump conditions, but does not change them in any
way. Equations \ref{equ:if1}, \ref{equ:if2} and \ref{equ:apf3} can be numerically integrated for any
cooling function resulting in an analytic profile from the specified
down-stream conditions to the shock front. The shock front is reached
when the adiabatic jump condition of equation \ref{equ:stop} is satisfied.

\section{Shock-tube}
\subsection{Description of the code}
The shock tube simulations were performed using a smoothed particle
hydrodynamics code. The code was that described by Couchman, Thomas \&
Pearce (1995) but with modified geometry to make the
simulation volume cuboidal with dimensions of $6\times 6 \times
100$. The shock front was initially set to be at z coordinate $50$
with gas flowing in the z direction. On what will be the upstream side
of this point, particles are placed one per unit volume and given unit
mass. On the downstream side, particles were placed four per unit
volume. Their mass however is allowed to vary in order for the gas to
have any density profile that is initially desired. The density
profile, along with the temperature and velocity of the gas, is read
in upon the start of a simulation. For this geometry the simulation
thus starts with $9000$ particles. The particles on each side of the
shock front were allowed to relax independently to a glass-like
initial state.

The sides of the simulation box were periodic, with the exception of
each end, where gas was fed into the box at the appropriate rate
upstream and removed from the simulation a sufficient distance
downstream so as not to effect any cooling tail. A padding region of
two mesh cells at each end was necessary and gas within this region
had its position updated on each time-step, but did not have its
energy, velocity or density updated. (The code searches for $32$
neighbours, thus needing a search radius of approximately $2$ units
when particles are located one per cell. It is this distance that
determines the size of the padding region as any particle closer than
this distance to the end of the simulation volume would experience a
non-symmetric force.) All simulations were performed in the rest frame
of the shock and the units the code used were dimensionless until a
cooling function was chosen.

The cooling function, and ionization/recombination rates were
integrated using RK4 from Press et al.~(1992),
this routine being called at the end of each time-step. All
the rates used were tabulated on the first timestep. 

\section{Stable In-shock cooling}
\subsection{Theory}
The cooling function for objects with high virial temperatures (above
$10^{4}$K) collapsing after the Universe has been enriched with metals
can be written as the sum of three terms: a bremsstrahlung term,
$\lambda_{1}$, a metal line-cooling term, $\lambda_{2}$ and a H-He
term, $\lambda_{3}$.
The fit for each term is split into two temperature regimes. For
$T>10^{5}$K we have, where $\lambda=\Lambda/$erg ${\rm cm}^{3}$ ${\rm s}^{-1}$:
\begin{equation}
\lambda_{1}=5.2\times10^{-28}\left(T/\K\right)^{\frac{1}{2}}
\\
\label{equ:cool1}
\end{equation}
\begin{equation}
\lambda_{2}=1.7\times10^{-18}\left(T/\K\right)^{-0.8}{\rm Z/Z_{\odot}}
\\
\end{equation}
\begin{equation}
\lambda_{3}=1.4\times10^{-18}\left(T/\K\right)^{-1}
\\
\end{equation}
For $10^{5}\K>T>10^{4}$K we have:
\begin{equation}
\lambda_{1}=0
\\
\end{equation}
\begin{equation}
\lambda_{2}=1.7\times10^{-27}\left(T/\K\right) {\rm Z/Z_{\odot}}
\\
\end{equation}
\begin{equation}
\lambda_{3}=1.4\times10^{-28}\left(T/\K\right)
\\
\label{equ:cool2}
\end{equation}
where ${\rm Z/Z_{\odot}}$ is the metallicity in terms of solar.

The above cooling function is a good approximation to within a factor
of $2$ to Raymond \& Smith (1977).
The nature of this cooling function is such that the cooling rate
rises abruptly as one moves from a temperature $T<10^{4}$K to
$T>10^{4}$K. The cooling rate peaks at $10^{5}$K and then decreases,
reaching a minimum at $2\times10^{7}$K. At this point bremsstrahlung
cooling starts to become effective and the cooling rate begins to
slowly rise as the temperature increases further.  
For gas being shocked from $10^{4}$K to a temperature of order $2\times10^{7}$K
the cooling function is of a suitable nature to produce 
a significant reduction in the cooling time of shocked gas, if in-shock
cooling is present. The cooling rate being most rapid at lower
temperatures and then decreasing as the temperature rises above
$10^{5}$K means that a modest reduction in the 
maximum temperature reached could make a significant difference to the
cooling time as it is the hottest gas that takes the longest to cool. 
Unfortunately a declining cooling function makes the shock unstable and
in this section we restrict ourselves to studying shocks
which do not reach the declining part of the cooling function, and
which thus settle down into a steady state.

The nature of such shocks can be specified by the ratio of the cooling
time to the shock
crossing time. If the cooling time is very long,
and the ratio is thus large, the effect of in-shock will be
negligible. Conversely, if the cooling time is very rapid, in-shock
cooling will tend to drive the shock towards an isothermal
transition. The cooling time itself will be determined by the physical
conditions that are being simulated and will be determined by the
density and temperature of the gas. The shock crossing time will
depend upon the mass resolution of the code and will vary as the cube
root of the particle mass.

\subsection{Methodology}
By specifying the downstream conditions of the shock, equations
\ref{equ:if1}, \ref{equ:if2} and \ref{equ:apf3} were numerically integrated for $\Lambda$ of section $4.1$, and a profile of the
cooling tail was obtained. The integration was stopped at the shock
front, when equation \ref{equ:stop} was satisfied. This profile was then read
directly into the shock-tube to create the initial conditions, with the shock front being placed at the
boundary between the low and high mass particles. Particles within the
region of the cooling tail had their temperature, velocity and mass
adjusted to recreate the analytic shock profile. The simulations were
then allowed to run until a steady state-state was reached.

The upstream gas was started with a temperature of $5.95\times
10^{4}$K and the analytic jump temperature was $1.84\times
10^{5}$K. The density of gas far downstream had a density twelve times
that of the upstream gas. Two simulations were ran, one with a ratio of the shock crossing time to the
cooling time of $8$, and the other with a ratio of $1.2$. 
\subsection{Results}
\begin{figure}
\begin{center}
\psfig{width=8.7cm,angle=270,file=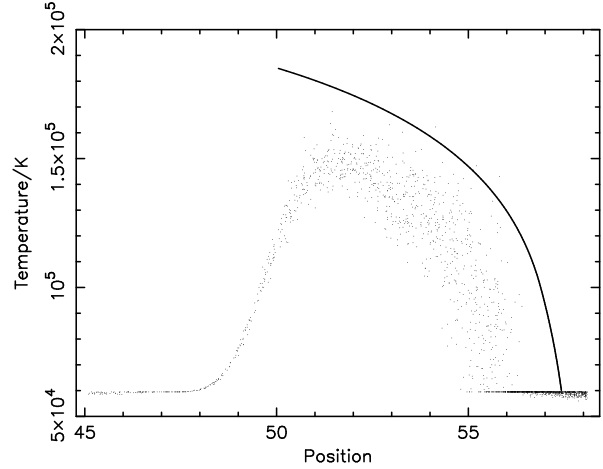}
\caption{Steady state temperature profile demonstrating in-shock
cooling. The solid line shows the analytic profile that would be
expected in the absence of in-shock cooling. The ratio of the
cooling-time to shock crossing time is of order $8$.}
\label{fig:stable1}
\end{center}
\end{figure}
\begin{figure}
\begin{center}
\psfig{width=8.7cm,angle=270,file=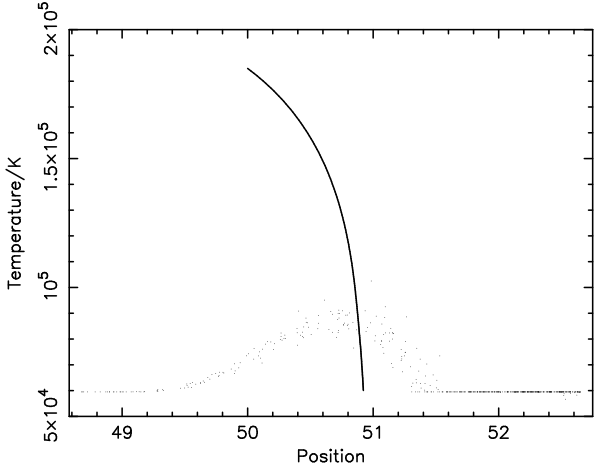}
\caption{Steady state temperature profile demonstrating in-shock
cooling. The solid line shows the analytic profile that would be
expected in the absence of in-shock cooling. The ratio of the
cooling-time to shock crossing time is of order $1.2$.}
\label{fig:stable2}
\end{center}
\end{figure}
Figures \ref{fig:stable1} and \ref{fig:stable2} show the temperature
profiles as a function of position for two shocks with ratios of $8$
and $1.2$ respectively. Despite the upper part of the theoretical
shock being in the unstable regime the simulations were found to
settle down into a steady state. This was aided by in-shock cooling
which reduces the maximum temperature reached. The solid lines in show
the analytic temperature profiles. We can see that the maximum
temperature obtained is approximately $25$ percent less than the
analytic jump temperature in Figure \ref{fig:stable1} , and $50$
percent less for the simulation of Figure \ref{fig:stable2} with a
particle mass $512$ times greater.

The effect that these reductions in jump temperature have on the
cooling tails can be seen to be somewhat erratic.  In both cases the
cooling time of the gas decreases by a factor of approximately $2$,
but the spatial extent of the cooling tail depends upon the
resolution.  The reason for this is that spatial location is not an
accurate indication of the cooling time, as the latter is dependent
upon the density. We would not necessarily expect the shock-tube
simulations to have the same density profile or the same density jump
at the shock front as the theoretical profile.  However they do
converge to the correct jump conditions far downstream.

The sort of objects in numerical simulations where in-shock cooling
will be important are those with a low ratio of cooling time to shock
crossing time and which cool on a time-scale of order the dynamical
time of the virialized halo.  Objects which cool on a time-scale
much longer than this are generally of little interest as they
will not collapse sufficiently quick to form luminous objects. At the
other extreme, if an object cools on a time-scale much less than the
dynamical time the gas will be in a state of free-fall
collapse; in-shock cooling will certainly be present in this case as
the cooling is so rapid, but the correct conditions will be recovered
on a time-scale less than the dynamical time.

As an example, consider the collapse of an object with a cooling time
equal to its collapse time, in a cosmological simulation with a
dark-matter particle mass of $3\times 10^{10}M_{\odot}$. Taking
$\Omega_{b}=0.06$, a galaxy with total mass of
$1.7\times10^{12}M_{\odot}$ collapsing at redshift $2$ would have a
shock velocity at about $350$km$\rm{s}^{-1}$. Taking the shock to be
spread over $3$ inter-particle spacings, we find the ratio of the
cooling time to the shock crossing time to be $1.2$. If the same
object were to be simulated with a particle mass $1000$ times lower,
the ratio would increase to $12$. The shock-tube simulations described
in this section suggest that, for either large-scale cosmological
simulations with a ratio of $1.2$, or simulations of a single galaxy
with a ratio of $8$ the effect of in-shock cooling will reduce the
maximum temperature reached by between $25$ to $50$ percent and reduce
the cooling time by a factor of approximately $2$.

\section{Simulations with electron cooling}
\subsection{Theory}
The most important cooling mechanisms for a primordial gas with a
temperature above $10^4$K result from recombination of electrons with
hydrogen ions and collisional excitation of neutral hydrogen by
collisions with free electrons, the excited hydrogen atom then emits a
photon which, should it not be re-absorbed, amounts to a reduction in
the internal energy of the gas. The cooling terms, taken from Haiman,
Thoul and Loeb (1996) are:
\begin{equation}
\frac{\Lambda}{{\rm cm}^{3}{\rm s}^{-1}}=7.5\times10^{-19}\left(1+T_{5}^{\frac{1}{2}}\right)^{-1}
{\rm exp}\left(\frac{-1.18348}{T_{5}}\right)
\\
\label{equ:emc}
\end{equation}
\begin{equation}
\frac{\Lambda}{{\rm cm}^{3}{\rm s}^{-1}}=4.02\times10^{-19}\left(\frac{T_{5}^{\frac{1}{2}}}{1+T_{5}^{\frac{1}{2}}}\right)
{\rm exp}\left(\frac{-1.57809}{T_{5}}\right)
\\
\label{equ:emc2}
\end{equation}
where $T_{5}$ is the temperature in units of $10^{5}$K.  The rate of
energy loss from the gas will be proportional to the number densities
of both neutral hydrogen and free electrons. The ionization level thus
needs to be known and is generally not equal to its equilibrium value
for the situations we shall consider. In order to trace the ionization
level we need to specify the ionization and recombination rates. For
the purpose of this section we consider a gas composed entirely of
hydrogen which is ionized through the reaction:
\begin{equation}
\Hyd + e^{-} \longmapsto \Hyd^{+} + 2e^{-}
\end{equation}
This proceeds at a rate (Black 1978):
\begin{equation}
\frac{\gamma_{{\rm i}}}{{\rm
cm}^{3}{\rm s}^{-1}}=1.85\times10^{-8}T_{5}^{\frac{1}{2}}{\rm exp}\left(\frac{-1.578091}{T_{5}}\right) 
\\
\label{ionize}
\end{equation}
The gas recombines via the reaction:
\begin{equation}
\Hyd^{+} + e^{-} \longmapsto \Hyd + h\nu
\\
\label{recomb}
\end{equation}
which proceeds at a rate (Hutchins 1976):
\begin{equation}
\frac{\gamma_{r}}{{\rm cm}^{3}{\rm
s}^{-1}}=1.19\times10^{-13}T_{5}^{-.64} 
\end{equation}
\subsection{Methodology}
For the electron-cooling runs, the gas was started with jump
conditions for a perfectly isothermal shock. The motivation for this
approach is that the cooling time is short, meaning that the perfectly
isothermal conditions are reasonably close to the stable profile and
should soon evolve to that state. The mass of the downstream particles
was adjusted so that the correct density jump was obtained and the
simulations were then allowed to run until the developing temperature,
density and ionization profiles reach a stable state. Gas upstream had
an ionization level of $4\times 10^{-4}$ and a temperature of
12\,000\,K, which was the minimum temperature permitted in the
simulations. Gas at or below 12\,000\,K did not have its ionization
level updated and was not allowed to cool further as this would slow
down the code too much. Instead, cooling below 12\,000\,K was estimated
analytically.

Run 1 is a fiducial run with units designed to represent an object at
a redshift of 20 containing 10$^6$ particles with a total baryonic
mass of $4\times 10^{8}M_{\odot} $ and a virial temperature of
78\,000\,K. Runs $2$ and $3$ have the same mass resolution as the
fiducial run but represent objects with higher and lower virial
temperatures respectively. Runs $4$ and $6$ have the same virial
temperature as the fiducial run but lower and higher mass resolution
respectively. Runs $5$ and $7$ have a lower virial temperature than
the fiducial but different mass resolutions. The mass resolution is
represented by the number of particles an object of mass
$4\times10^8M_{\odot}$ would contain, although the simulations which
have been performed have the same number of particles in every case.

\subsection{Results}
Table $1$ below lists the results of seven shock-tube simulations
using the cooling functions of equations \ref{equ:emc} and
\ref{equ:emc2}.  From left to right, the column headings of Table~1
are: run number; the number of particles an object of mass $4\times
10^{8}M_{\odot}$ would contain, $N_{\rm obj}$; the (analytic) virial
temperature $T_a$ (the temperature the shocked gas should jump to);
the ratio of the maximum temperature obtained in the simulations to
the virial temperature, $T_s/T_a$; the analytic ionization level for
gas cooling from the virial temperature to 12\,000\,K, $x_a$; the
ionization level of gas in the simulation that has cooled back to
12\,000\,K divided by the analytic level, $x_s/x_a$; the ratio of
the simulated to analytic cooling times from the peak of the shock,
down to 12\,000\,K, $\tau_s/\tau_a$; the ratio of the numerical
to the analytic cooling times to reach 600\,K, $\tau_{sf}/\tau_{af}$;
and the ratio of the numerical to the analytic \Htwo\ fractions at
600\,K, H$_{2,s}$/H$_{2,a}$.

The analytic ionization level at 12\,000\,K, $x_a$, assumes an
adiabatic shock followed by rapid cooling and was calculated by
integrating the shock conservation equations \ref{equ:if1},
\ref{equ:if2} and \ref{equ:apf3} with the cooling functions of
equations \ref{equ:emc} and \ref{equ:emc2} using non-equilibrium
chemistry.  Cooling below 12\,000\,K was calculated allowing the gas
to cool at a constant density to $600$K via the \Htwo\ cooling
mechanism.  A detailed treatment of the gas chemistry was used
incorporating formation and destruction of \Htwo, ionization and
recombination, the cooling functions of equations \ref{equ:emc} and
\ref{equ:emc2} and the \Htwo\ cooling function. The destruction and
cooling rates of \Htwo\ was taken from Martin, Schwarz and Mandy
(1996).  The simulated cooling time from 12\,000\,K to 600\,K,
$\tau_{sf}$, and \Htwo\ fraction H$_{2s}$ use $x_s$ as the starting
ionization level; $\tau_{af}$ and H$_{2a}$ use $x_a$ as the input
ionization.

\begin{table*}
\begin{tabular}{ccrclccrc}   \hline\hline
Run No. & $N_{\rm obj}$ & \hfil$T_a$/K\hfil & $T_s/T_a$ & \hfil$x_a$\hfil & 
\hfil$x_s/x_a$\hfil & $\tau_{sf}/\tau_{af}$ & $\tau_{s}/\tau_{a}$ &
H$_{2,s}$/H$_{2,a}$\\ \hline 
$1$ & $10^{6}$ &  $78\,000$ & $0.27$ &
$0.14$ & 0.11 & $236$ & $2.14$\ \mbox{}& $0.50$ \\
$2$ & $10^{6}$ &  $168\,000$ & $0.15$ &
$0.58$ & 0.08 & $512$ & $2.01$\ \mbox{}& $0.51$ \\
$3$ & $10^{6}$ &  $39\,000$ & $0.45$ &
$0.029$ & 0.24 & $48$ & $1.74$\ \mbox{}& $0.61$ \\
$4$ & $10^{3}$ &  $78\,000$ & $0.22$ &
$0.14$ & 0.09 & $941$ & $2.40$\ \mbox{}& $0.45$ \\
$5$ & $10^{3}$ &  $39\,000$ & $0.44$ &
$0.029$ & 0.16 & $139$ & $2.10$\ \mbox{}& $0.52$ \\
$6$ & $10^{9}$ &  $78\,000$ & $0.39$ &
$0.14$ & 0.24 & $42$ & $1.63$\ \mbox{}& $0.64$ \\
$7$ & $10^{9}$ &  $39\,000$ & $0.81$ &
$0.029$ & 0.41 & $9$ & $1.40$\ \mbox{}& $0.74$ \\ \hline\hline
\end{tabular}
\caption{Steady state properties of shock tube simulations
with cooling due to collisional excitation of free electrons with
neutral hydrogen.}
\label{tab:t1}
\end{table*}
\subsection{Discussion}
\subsubsection{Fiducial Run}
The effect of in-shock cooling is dramatic and striking in the
fiducial run. The maximum temperature the shock reaches is only $0.27$
of its theoretical value. This leads to an increase in the cooling
time for the gas to cool back down to 12\,000\,K by an astonishing
factor of $236$. There are a number of contributing factors to this
extreme figure. Firstly, although the ionization level at 12\,000\,K
is $9$ times lower in the shocktube than expected, this is a
conservative estimate of the difference, as the factor is somewhat
higher than this at other points on the cooling tail. This is because
the lower jump temperature obtained results in a much slower rate of
of ionization. Also, the density of the theoretical profile is
somewhat higher at any given temperature (above 12\,000\,K) as the gas
has reduced its temperature by a greater factor than gas at the same
temperature in the shock tube. e.g. for the fiducial run, at
22\,000\,K the theoretical profile has a density of $16$ while gas at
this temperature in the shocktube has has a density of only
$3.5$. These effects combined produce the high factors of
$\tau_s/\tau_a$ seen in Table~1. The consequences of this are of
extreme importance as simulations of such objects will be meaningless
unless in-shock cooling can be removed, as the increase in the cooling
times of such objects will slow the collapse.

\begin{figure}
\begin{center}
\psfig{width=8.7cm,angle=270,file=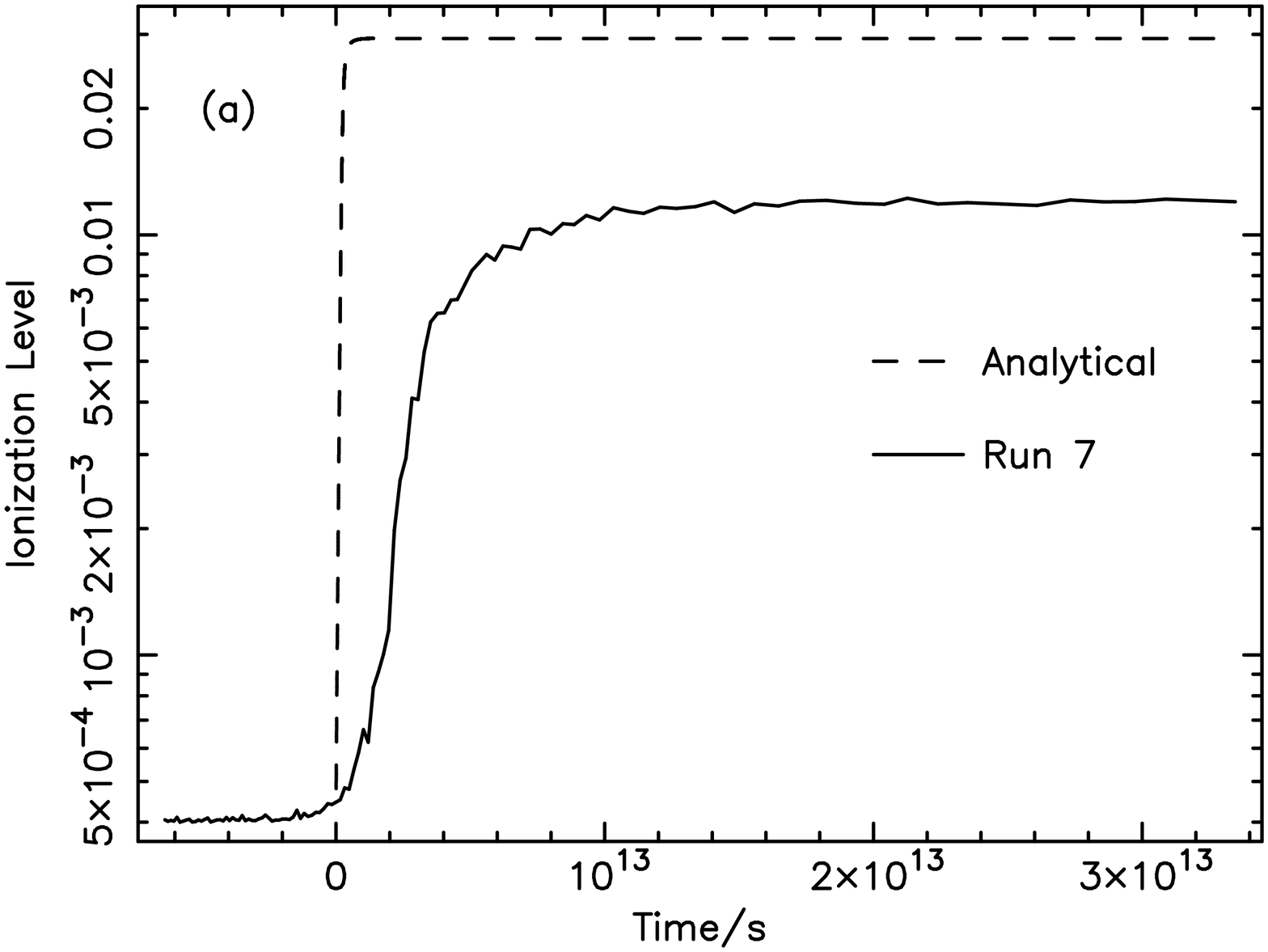}
\psfig{width=8.7cm,angle=270,file=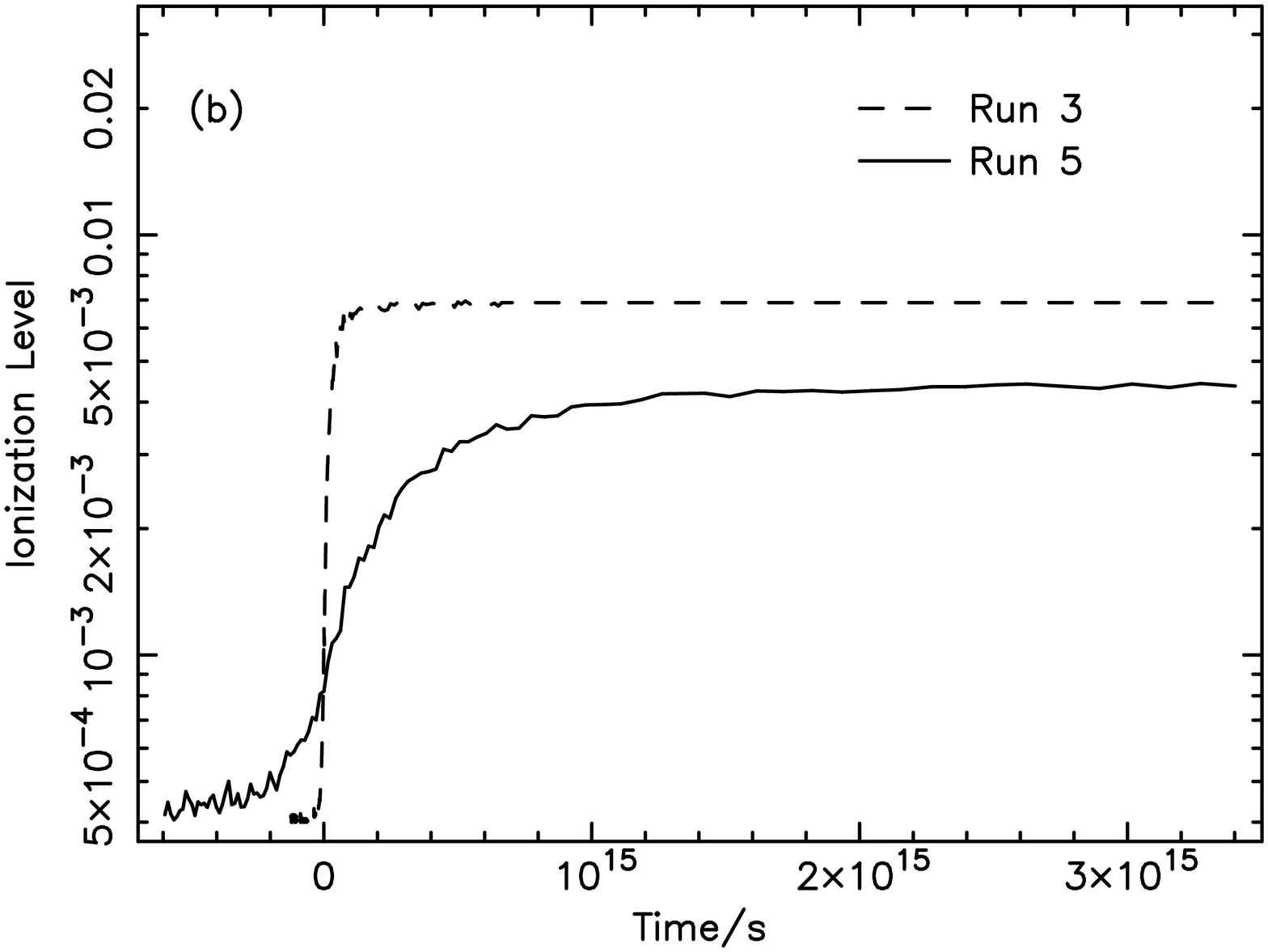}
\caption{(a) Variation of Ionization Level with time for gas passing
through the shock of the fiducial run (dashed line) and run 7 (solid
line). (b) Variation of Ionization Level with time for gas passing
through the shock of run 3 (dashed line) and run 5 (solid line).}
\label{fidprofs}
\end{center}
\end{figure}

Also of importance is the ability of an object to cool to temperatures
below 12\,000\,K when molecular hydrogen becomes the dominant coolant.   
Molecular hydrogen is formed via the reactions
\[
\Hyd + e^{-} \longmapsto \Hyd^{-} + h\nu
\\
\]
\[
\Hyd^{-}+ \Hyd \longmapsto \Hyd_{2} + e^{-}
\]
Its production is thus proportional to the ionization level, the free
electrons acting as a catalyst.  The effect of in-shock cooling on the
fiducial run was found to halve the \Htwo\ abundance and to increase
the cooling time from 12\,000\,K to 600\,K by a factor of two.

\subsubsection{Effect of Virial Temperature}
Increasing the shock jump temperature, whilst retaining the same
spatial and mass resolution results in a modest increase in the
maximum temperature obtained at the shock front. This difference is
sufficient to cause the ionization level of the cooled downstream gas,
$x_{s}$ to be higher than that obtained in the fiducial run by a
factor of almost three. However, the correct ionization level is still
some way from being obtained in each of runs $1$, $2$ and
$3$. Comparing these runs shows that the fractional temperature
reached increases with decreasing virial temperature, and the
ionization level gets closer to its theoretical value. However, these
trends do not appear to carry over to the cooling times and \Htwo\
abundances when the virial temperature increases from 78\,000\,K to
16\,8000\,K. This is due to the complex interaction between reaction
rates and cooling times.

\subsubsection{Effect of resolution}
As one might expect, improving the mass resolution reduces the
severity of any in-shock effects. For jump temperatures of 78\,000\,K
and 39\,000\,K, as the mass resolution is improved there is a steady
convergence towards the correct values of the \Htwo\ abundance, the
cooling time to $600$K, the cooling time from the virial temperature
to 12\,000\,K and the jump temperature. Essentially the convergence of
the first three is determined by the convergence towards the correct
jump temperature, as it is this temperature that determines the extent
to which the gas is re-ionized which in turn determines the rate of
\Htwo\ production and the cooling time. Figure \ref{fidprofs} shows
the ionization level as a function of time for an analytical shock and
for three shocks each with a virial temperature of 39\,000\,K, but with
different mass resolutions.  In each case we have defined time zero to
be the time when gas reaches the maximum temperature. Negative time
represents gas which is being heated by the shock but has not yet
reached the peak.  These figures show the effect of lowering the
resolution from the analytical profile, through runs 7, 3 and
5. Note that the y-scale is the same in each panel but the x-scale
is not.

We can see from the graphs that as resolution is improved
there is a convergence of the final ionization level and cooling
time towards the theoretical profile.  However, even if an object
were to contain $10^{9}$ particles the effects of in-shock cooling
would still remain significant, most noticeably for $\tau_{s}/\tau_{a}$. Such resolutions are at present not
possible, and even if they were the quantities we have considered are yet to converge.
We must conclude that the resolution of typical simulations being
performed is hopelessly inadequate and if sensible results are to be
obtained a way of removing in-shock cooling must be found.

\section{Temperature oscillations of unstable shocks.}
\subsection{Theory}
For a planar radiative shock with a cooling function $\Lambda \propto
T^{\alpha}$, Chevalier and Imamura (1982) showed that for $\alpha \leq
0.4$ the shock is unstable
against perturbations and will undergo oscillations both spatially and
in the jump temperature. The latter results from the fact that as the
shock front moves, the relative velocity of the upstream gas in the
frame of the shock varies. This in turn results in a different
solution of the jump equations. The situation is similar for stable
shocks also: if perturbed, they also undergo oscillations but with a
decaying amplitude. 
For values of $\alpha$ between $-1$ and $2$ Chevalier and Imamura
found the
frequency of oscillation, $\omega$, to vary 
between $0.26$ and $0.31$ with units of $u_{in}/\bar{x}_{s}$ where
$\bar{x}_{s}$ is the mean distance from the shock front to the point
upstream where the gas has cooled to its upstream temperature, and $u_{in}$ is
the velocity of the upstream gas.
The maximum displacement of the shock front from its mean position, $x_{max}$,
was also calculated and was found to be $x_{max}=\tau_{cool}v_{s}$,
where $\tau_{cool}$ is the steady state cooling time and $v_{s}$ the
speed the shock front is moving at.    

\subsection{Methodology}
The following simulations verify the instabilities found by Chevalier
and Imamura (1982) for the case of a realistic cooling function composed
of a number of power law terms, (some of which are stable).
An analytic profile of the cooling tail was calculated as in Section $4.2$.
The gas upstream had a temperature of $50000$K with a density
jump of $100$ for the downstream gas. This results in a temperature
jump of a factor of $19.7$ to $9.8\times 10^{5}$K. The cooling function used was
that of Section $4$, meaning that gas with a temperature above
$10^{5}$K is in the unstable regime. The fact that the shock is
unstable means that a steady state will never be reached so the
simulation was ran until the nature of the oscillations became apparent.   
\begin{figure}
\begin{center}
\psfig{width=8.7cm,angle=270,file=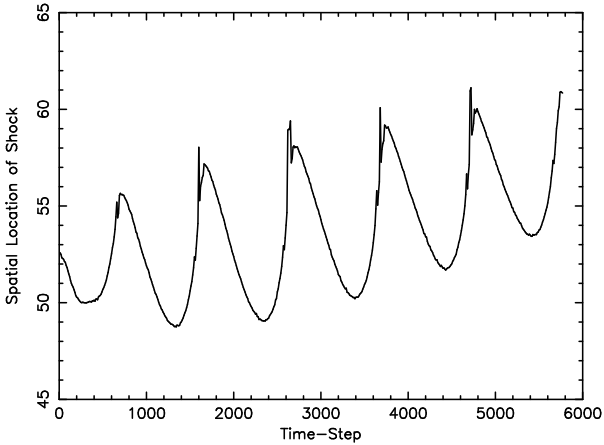}
\caption{Position of shock front as a function of time for an unstable
radiative shock.}
\label{usp}
\end{center}
\end{figure}

\begin{figure}
\begin{center}
\psfig{width=8.7cm,angle=270,file=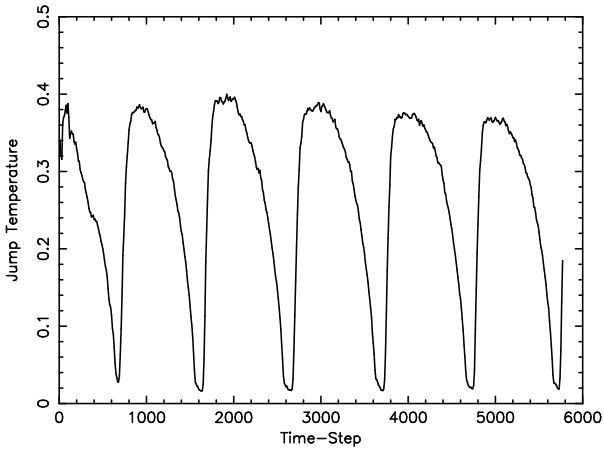}
\caption{Oscillations of the jump temperature of an unstable shock as
a function of time.}
\label{ust}
\end{center}
\end{figure}
\subsection{Results}
The unstable nature of both the position and jump temperature of radiative
shocks is well 
demonstrated upon consideration of Figures \ref{usp} and \ref{ust}. Although the cooling
function used is not strictly a power law, but
rather a combination of power laws, we find that both the magnitude of
the spatial oscillations, and their period, are in reasonable agreement with
the work of Chevalier and Imamura as presented above. The
frequency of oscillations $\omega$ was found
to be $\omega=0.31$, where
$\omega=\left(2\pi/P\right)\left(\bar{x}_{s}/u_{in}\right)$ and P is the observed period of oscillations.
The majority of heated gas
in the simulation has $\alpha\sim-0.8$. 
For the case of $\alpha=-1$, Chevalier and Imamura found the
oscillation frequency to be $\omega=0.26$.
The agreement is
reasonable, although interpolation of the Chevalier and Imamura values
to intermediate values of $\omega$ is somewhat uncertain as
there is no obvious linear trend in their results.  
The maximum displacement of the shock was found to be
$x_{max}=8.0$ which agrees well with the theoretical value of
$x_{max}=7.8$.

The effects of in-shock cooling on the simulations is uncertain but,
given that we know it is present, it may account for the extreme
oscillations down to an isothermal transition as seen in Figure 5.  We
note, however, that the period and spatial magnitude of the oscillations
appears to be correct even when in-shock cooling is present. 
In the absence of in-shock cooling, any halo with a virial temperature
of more than a few times 10$^5$K (i.e.~galaxies) will be subject to
such an instability.

\section{Conclusion}
In this paper we have clearly demonstrated that in-shock cooling has
important consequences for numerical simulations of galaxy formation,
especially at
high redshifts, where collisional excitation of hydrogen by free
electrons, cooling by recombination and cooling by molecular hydrogen are the relevant
coolants, but also for galaxies with higher virial temperatures
collapsing at lower redshifts.

The effect of in-shock cooling on a stable radiative shock was found to
decrease the maximum temperature reached by $25$ percent when the
ratio of the cooling-time to shock crossing time was $8$. This value
increased to $50$ percent for a ratio of $1.2$. The result of this was
that the profiles of the cooling tails were found to have different
spatial extents from the theoretical profile.
From this we conclude that although the effects of
in-shock cooling are quite small, the collapse rate of such objects
could be altered along with the spatial extent of any cooling tail.

The effect on primordial objects of in-shock cooling was found to be
extremely significant: the ionization level emerging from the shock
was significantly lower by a factor of $2.4$ in the best resolved
case, and up to a factor of $12.1$ in the worst case. The implications
for the cooling time of the gas ranged from an increase of a factor of
$9$ up to nearly $1000$, for the cooling times above 12\,000\,K. Below
12\,000\,K the result was less extreme, ranging from a factor of $1.4$
to $2.4$. This difference is nevertheless important and will
significantly slow the collapse of any object. The resulting molecular
hydrogen abundance of the gas after it was allowed to cool to $600$K
was found in all cases to decrease from its analytic value by a factor
of around 2.

For the case of an unstable shock we find oscillations of the spatial
location of the shock front and the maximum temperature reached. It is
likely that the extreme oscillations of temperature observed (cooling
to a virtually isothermal state) are fueled by in-shock cooling.  For
the resolutions currently achievable, an incorrect result will be
obtained for simulations of galaxy formation.

\begin{figure}
\begin{center}
\psfig{width=8.7cm,angle=270,file=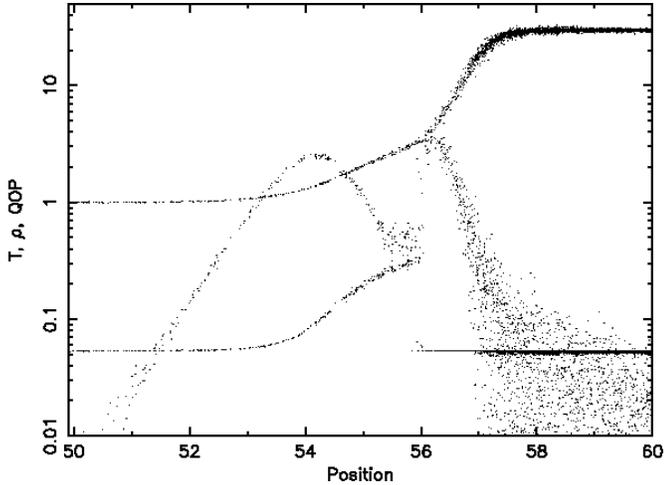}
\caption{Graph showing the properties of temperature, density and the
ratio of the artificial pressure to the pressure for a shock where
in-shock cooling has been removed. The temperature starts at $0.054$
and is heated to $0.30$ at position $56$, this temperature being the
correct jump temperature. The gas then cools extremely rapidly back
down to its upstream value. The density starts at $1$ and increases as
the gas shocks and the cools until it reaches its downstream value of
$12$. The ratio $q/p$ has a double peaked structure, each peak being
caused by shock heating and rapid cooling respectively.}
\label{fig:qopfig}
\end{center}
\end{figure}

Finally, Figure \ref{fig:qopfig} shows a successful attempt to remove
in-shock cooling from the fiducial run. The method used to remove the
effect was case-specific and was achieved by allowing the gas to cool
only when its temperature is in excess of the adiabatic jump temperature or
its density exceeds the adiabatic jump density. The upper curve in
Figure \ref{fig:qopfig} shows the gas density, the lower curve shows
the temperature and the middle curve, with the double maximum, shows
the dimensionless ratio of the artificial pressure to the pressure,
$q/p$. The density and temperature are plotted in code units. The finite
width of the shock can clearly be seen and the correct jump conditions
and ionization level both at the peak of the shock and downstream
are recovered. The correct cooling tail down to a temperature of
12\,000\,K is also recovered.

The importance of the ratio $q/p$ is that it is a dimensionless
indicator of when the gas is being shocked and might therefore be used
as a method of removing in-shock cooling from 
other simulations, by means of preventing the gas from cooling when
$q/p$ exceeds a certain threshold value (i.e. when the gas is being
shocked). However, this approach is unlikely to prove successful for
two reasons. Firstly, consideration of the double maximum structure of
$q/p$ as seen in Figure \ref{fig:qopfig} means than gas which is
shocking does not necessarily have a value of $q/p$ greater than gas
which is not. Indeed, gas which is radiating rapidly (which is the
case for gas whose density is increasing) has values of $q/p$ as high
as gas which is being shocked. Secondly, although $q/p$ is
dimensionless and in theory could produce a threshold value true for
all scales of simulations, we would not expect the threshold value to
be the same for shocks of different strengths. These two reasons rule
out the use of $q/p$ as a possible means of eliminating in-shock
cooling. This was tested in a large-scale cosmological
simulation. Different objects were found to have very different values
of $q/p$ and it was obvious that there was no single value that could
be used to determine when gas was being shocked.     

\section{Acknowledgments}
The authors would like to thank Paul Nulsen for discussion about the
effects of in-shock cooling.

\end{document}